\documentclass{article}
\usepackage{spconf,amsmath,graphicx,subfigure}
\usepackage{arydshln}
\usepackage{multirow}
\usepackage{booktabs}
\usepackage{color}
\usepackage{amsfonts}
\usepackage{longtable}
\usepackage{pifont}
\usepackage{hyperref}
\usepackage{cite}

\DeclareUnicodeCharacter{0301}{\'}

\title{Codec-SUPERB @ SLT 2024: A lightweight benchmark for \\ neural audio codec models}

\name{\begin{tabular}{c}
    Haibin Wu$^{1}$, Xuanjun Chen$^{1\dagger}$, Yi-Cheng Lin$^{1\dagger}$, Kaiwei Chang$^{1\dagger}$, Jiawei Du$^{1\dagger}$, Ke-Han Lu$^{1\dagger}$, \\
    Alexander H. Liu$^{2\dagger}$,
    Ho-Lam Chung$^{1\dagger}$, Yuan-Kuei Wu$^{1\dagger}$, Dongchao Yang$^{5\dagger}$, Songxiang Liu$^5$, \\ Yi-Chiao Wu$^4$, 
    Xu Tan$^3$, James Glass$^2$, Shinji Watanabe$^6$, Hung-yi Lee$^{1}$
\end{tabular}
\thanks{$^\dagger$ Equal contribution}
}
\address{$^1$National Taiwan University, $^2$Massachusetts Institute of Technology, $^3$Microsoft Corporation, \\ $^4$Meta, $^5$The Chinese University of Hong Kong, $^{6}$ Carnegie Mellon University}

\begin{document}
%
\maketitle
\begin{abstract}
Neural audio codec models are becoming increasingly important as they serve as tokenizers for audio, enabling efficient transmission or facilitating speech language modeling.
The ideal neural audio codec should maintain content, paralinguistics, speaker characteristics, and audio information even at low bitrates.
Recently, numerous advanced neural codec models have been proposed.
However, codec models are often tested under varying experimental conditions.
As a result, we introduce the Codec-SUPERB challenge at SLT 2024 \footnote{https://codecsuperb.github.io/}, designed to facilitate fair and lightweight comparisons among existing codec models and inspire advancements in the field. 
This challenge brings together representative speech applications and objective metrics, and carefully selects license-free datasets, sampling them into small sets to reduce evaluation computation costs.
This paper presents the challenge's rules, datasets, five participant systems, results, and findings.
\end{abstract}

\section{Introduction}
\label{sec:intro}

Neural audio codecs were originally created to compress audio data into compact codes, for better transmission and storage \cite{pohlmann2000principles}.
Recently, these codec models have gained significant interests because they can bridge audio and language processing.
Researchers are exploring their use as tokenizers, which convert continuous audio into discrete codes that can be used to develop audio language models (LMs) \cite{borsos2023audiolm,wang2023neural,kreuk2022audiogen,agostinelli2023musiclm,wu2024towards}.
The dual roles of neural audio codecs—reducing data transmission time and acting as tokenizers—emphasize their importance. 
In recent years, there have been notable advancements in codec models \cite{defossez2022high, zeghidour2021soundstream, borsos2023soundstorm, wu2023audiodec, yang2023hifi, du2023funcodec, zhang2023speechtokenizer, kumar2023high, ju2024naturalspeech, ji2024language, wu2024scoredec,zheng2024srcodec, zheng2024supercodec, xu2024lightcodec,yang2024generative,liu2024semanticodec,ai2024apcodec,siuzdak2023vocos,li2024single,guo2024addressing,yang2024uniaudio} \footnote{https://github.com/ga642381/speech-trident}, and \cite{wu2024towards} conducts a brief overview about codec models and speech LMs. 
The ideal neural audio codec should maintain content, paralinguistics, speaker characteristics, and audio information even at low bitrates. 
However, the optimal codec model for audio information preservation remains unclear, as various neural codec models are evaluated under their specific experimental conditions. 

Chang et al. \cite{chang2024interspeech} and Mousavi et al. \cite{mousavi2024dasb} compared various types of discrete audio tokenizers by training downstream generative and discriminative models based on the extracted discrete audio tokens.
Chang et al. concentrated on automatic speech recognition, text-to-speech, and singing voice conversion. 
In contrast, Mousavi et al. explored a range of discriminative tasks, including speech recognition, keyword spotting, intent classification, speaker recognition and emotion recognition, along with generative tasks such as text to speech, speech separation and speech enhancement.
The evaluation pipeline of our challenge is training-free and designed to help codec developers quickly obtain preliminary results, serving as a reference for their further development.
A previous training-free work \cite{wu2024codec} provided a wide analysis of the resynthesized audio quality from different codec models. 
However, \cite{wu2024codec} mentioned that their evaluation required significant computation time and computation resources. 
This challenge improved the evaluation pipeline from \cite{wu2024codec} by replacing all license-restricted datasets with license-free ones and reducing the size of the evaluation data. 
This created a lightweight benchmark that makes evaluating different codec models more efficient and easier.
The challenge's evaluation pipeline offers the advantages of being training-free, license-free, lightweight, and computationally efficient.



\section{Challenge overview}

This challenge will comprehensively analyze the quality of audio resynthesized by various codec models from both application and signal perspectives \cite{wu2024codec}.
Various codec models will be used to resynthesize the audio, and the quality of the resynthesized audio will be evaluated using application-level metrics (as detailed in Section~\ref{sec:app}) and signal-level metrics (as detailed in Section~\ref{sec:obj}).
We prepare an easy-to-follow script for participants, which includes open dataset download, environment installment, and evaluation.

\begin{table}[t]
\small
\centering
\caption{Dataset information. \textbf{app} implies the dataset is used in application-level evaluation. \textbf{obj} implies the dataset is used in objective metrics evaluation.}
\vspace{3pt}
\resizebox{0.5\textwidth}{!}{%
\begin{tabular}{llll}
\toprule
\textbf{Speech dataset} & \textbf{Features} & \textbf{app} & \textbf{obj}\\
\midrule
Librispeech \cite{DBLP:conf/icassp/PanayotovCPK15} & diverse speaker, read audiobooks & \ding{51} & \ding{51}\\ 
VoxCeleb1 \cite{DBLP:conf/interspeech/NagraniCZ17} & diverse speaker, celebrities on YouTube & \ding{51} & \ding{51}\\ 
QUESST \cite{anguera2015quesst2014} & multi-lingual, low resource language & &\ding{51}\\
VoxLingua107 Top 10 \cite{valk2021slt} & multi-lingual, YouTube content & & \ding{51}\\
Fluent Speech Commands \cite{fluent} & spoken keyword commands & & \ding{51}\\
Audio SNIPS \cite{DBLP:conf/icassp/LaiCL0G21} & spoken commands, crowdsourced & & \ding{51}\\
CREMA-D \cite{6849440} & affective speech & & \ding{51}\\
RAVDESS \cite{livingstone2018ryerson} & affective speech & \ding{51} & \\
Libri2Mix \cite{cosentino2020librimix} & multi-speaker scenarios & & \ding{51}\\
\midrule
\textbf{Audio dataset}& \textbf{Features} & & \\
\midrule
ESC-50 \cite{piczak2015dataset} & diverse audio source & \ding{51} & \ding{51}\\
FSD-50K \cite{fonseca2022FSD50K} & diverse audio source & & \ding{51}\\
Gunshot Triangulation \cite{cooper2020gunshots} & diverse audio source & & \ding{51}\\
\bottomrule
\end{tabular}
}
\label{tab:dataset_info}
\end{table}

\subsection{Application}
\label{sec:app}
The application angle evaluation will comprehensively compare each codec's ability to preserve crucial audio information. 
This includes content (measured by word error rate (WER) for automatic speech recognition (ASR)), speaker timbre (measured by equal error rate (EER) for automatic speaker verification (ASV)), emotion (measured by accuracy for speech emotion recognition), and general audio characteristics (measured by accuracy for audio event classification).

\subsubsection{Automatic speech recognition (ASR)}


For the ASR evaluation, we use the Whisper model \cite{radford2022robust} to assess how well various codecs preserve context information within speech. The primary metric is word error rate. This evaluation is conducted on the LibriSpeech dataset \cite{DBLP:conf/icassp/PanayotovCPK15}, specifically focusing on the test-clean and test-other subsets, with a total random sample of 500 samples from both subsets. 

\subsubsection{Automatic speaker verification (ASV)}
Speaker information represents a unique aspect of speech. To assess the degree of speaker information loss in the resynthesized speech generated by neural codecs, we employ automatic speaker verification. We use the cutting-edge speaker verification model, ECAPA-TDNN \cite{desplanques2020ecapa} \footnote{https://github.com/TaoRuijie/ECAPA-TDNN}, as the pre-trained ASV model. The evaluation is performed on the Voxceleb test-O set \cite{DBLP:conf/interspeech/NagraniCZ17}, using equal error rate (EER) as the metric. EER provides a balance between false acceptances and false rejections.

\subsubsection{Emotion recognition (ER)}

In addition to speaker information, speech conveys emotional information. We employ speech emotion recognition to quantify the degree of emotional information loss due to speech resynthesis by codec models. For this evaluation, we utilize the emotion2vec model \cite{ma2023emotion2vec} \footnote{https://github.com/ddlBoJack/emotion2vec} on the well-known license-free emotion dataset, RAVDESS \cite{livingstone2018ryerson}.

\subsubsection{Audio event classification (AEC)}

We adopt the AEC task to evaluate how effectively different codecs preserve audio event information. This involves using a pre-trained AEC model to classify audio events in the re-synthesized audio. Specifically, we utilize the pre-trained Contrastive Language-Audio Pretraining (CLAP) model \cite{CLAP2023, CLAP2022}\footnote{https://github.com/microsoft/CLAP} for testing on the ESC-50 dataset \cite{piczak2015dataset}.

\subsection{Objective metrics}
\label{sec:obj}
The diverse set of signal-level metrics, including Perceptual Evaluation of Speech Quality (PESQ) \cite{rix2001perceptual}\footnote{We use the implementation from \cite{miao_wang_2022_6549559}}, Short-Time Objective Intelligibility (STOI) \cite{taal2010short}\footnote{https://github.com/mpariente/pystoi}, Signal-to-distortion ratio (SDR), Mel Spectrogram Loss (MelLoss) \cite{kumar2023high} \footnote{https://github.com/descriptinc/descript-audio-codec/tree/main}, enable us to conduct a complete evaluation of audio quality across various dimensions, encompassing spectral fidelity, temporal dynamics, perceptual clarity, and intelligibility. 

\subsection{Dataset}

To facilitate the development of codec techniques and ensure fair comparisons among challenge submissions, we have curated two datasets: the open set and the hidden set. The hidden set will remain undisclosed to participants throughout the challenge. The open set functions as the development set, allowing participants to evaluate and develop their models.

\subsubsection{Open set}
Below, we present the open sets utilized in this challenge. To address licensing concerns, certain datasets from the previous paper \cite{wu2024codec} were replaced or excluded. Details of the selected datasets can be found in Table~\ref{tab:dataset_info}. 
Additionally, we conducted random sampling of the data to reduce the size of the evaluation dataset, thereby accelerating the evaluation process and minimizing evaluation efforts.

\noindent\textbf{QUESST} 2014 dataset \cite{anguera2015quesst2014} comprises 23 hours of spoken documents in six under-resourced languages. The recordings are encoded at 8 KHz with 16-bit resolution, featuring diverse speech types and acoustic environments.

\noindent\textbf{Fluent Speech Commands} dataset \cite{fluent} includes 30,043 spoken utterances from 97 individuals, recorded as single-channel .wav files at a 16 kHz sampling rate. Each file contains a unique utterance designed for controlling smart-home devices or interacting with a virtual assistant, such as ``turn off the light in the bedroom". We utilize the test set for codec evaluation.

\noindent\textbf{LibriSpeech} \cite{DBLP:conf/icassp/PanayotovCPK15} is a widely used corpus of English speech data, containing approximately 1000 hours of audio recordings. The recordings feature a reading style, comprising utterances read from audiobooks. 
The test-clean and test-other sets are adopted for codec evaluation.

\noindent\textbf{Audio SNIPS}  \cite{DBLP:conf/icassp/LaiCL0G21} employs a text-to-speech (TTS) system to synthesize utterances from the SNIPS dataset, incorporating various speakers and accents. This dataset is tailored for concurrent speech recognition and natural language understanding tasks. We employ the test and validation splits for codec evaluation.

\begin{table}[htbp]
\centering
\small
\caption{Codec information. ``A" refers to the FunCodec \cite{du2023funcodec}. ``B$\sim$" refers to the SemantiCodec \cite{liu2024semanticodec}. ``C$\sim$" refers to the APCodec \cite{ai2024apcodec}. ``D$\sim$" refers to the AFACodec. ``E" refers to the SpeechTokenizer \cite{zhang2023speechtokenizer}.}
\begin{tabular}{llccc}
\toprule
\textbf{Codec} & \textbf{Bitrate} & \textbf{Parameter Num} & \textbf{Sampling Rate} \\
\midrule
A & 8 kbps & 57.83 M & 16k \\
\midrule
B1 & 0.34 kbps & 187.77 M & 16k \\
B2 & 0.35 kbps & 187.77 M & 16k \\
B3 & 0.68 kbps & 921.72 M & 16k \\
B4 & 0.70 kbps & 921.72 M & 16k \\
B5 & 1.35 kbps & 507.42 M & 16k \\
B6 & 1.40 kbps & 507.42 M & 16k \\
\midrule
C1 & 2 kbps & 69 M & 16k \\
C2 & 4 kbps & 69 M & 16k \\
\midrule
D1 & 2 kbps & 73.07 M & 16k \\
D2 & 7 kbps & 73.07 M & 44k \\
D3 & 7.5 kbps & 73.07 M & 48k \\
\midrule
E & 4 kbps & 103 M & 16k \\
\midrule
\bottomrule
\end{tabular}
\label{tab:codec_info}
\end{table}

\begin{table*}[h]
\centering
\caption{Comparison between codec models for the open set. ``None" means that no codec has been applied.}
\resizebox{\textwidth}{!}{
\begin{tabular}{llcccccccccccc}
\toprule
\multirow{2}{*}{\textbf{Codec}} & \multirow{2}{*}{\textbf{Bitrate}} & \multicolumn{4}{c}{\textbf{Application}} & \multicolumn{4}{c}{\textbf{Speech Signal-Level Metrics}} & \multicolumn{2}{c}{\textbf{Audio Signal-Level Metrics}} \\
\cmidrule(lr){3-6} \cmidrule(lr){7-10} \cmidrule(lr){11-12}
 & \textbf{(kbps)} & \textbf{WER (\%)} & \textbf{EER (\%)} & \textbf{ACC (\%)} & \textbf{ACC (\%)} & \multirow{2}{*}{\textbf{PESQ}} & \multirow{2}{*}{\textbf{STOI}} & \multirow{2}{*}{\textbf{SDR}} & \textbf{Mel} & \multirow{2}{*}{\textbf{SDR}} & \textbf{Mel} \\
 &  & \textbf{$\downarrow$ (ASR)} & \textbf{$\downarrow$ (ASV)} & \textbf{$\uparrow$ (ER)} & \textbf{$\uparrow$ (AEC)} &  &  &  & \textbf{Loss} &  & \textbf{Loss} \\
\midrule
None & - & 2.89 & 0.96 & 76.76 & 93.85 & - & - & - & - & - & - \\
\midrule
A & 8 & 3.13 & 1.56 & 75.21 & 83.30 & 2.63 & 0.93 & 6.85 & 1.86 & 0.11 & 2.18 \\
\midrule
B1 & 0.34 & 35.79 & 13.70 & 61.53 & 71.55 & 1.33 & 0.69 & -10.17 & 1.11 & -14.91 & 1.59 \\
B2 & 0.35 & 34.24 & 13.39 & 59.51 & 70.45 & 1.33 & 0.69 & -10.03 & 1.11 & -15.06 & 1.59 \\
B3 & 0.68 & 9.55 & 6.16 & 68.12 & 76.55 & 1.55 & 0.76 & -9.19 & 0.93 & -14.60 & 1.52 \\
B4 & 0.70 & 9.69 & 6.01 & 67.15 & 75.10 & 1.56 & 0.77 & -9.17 & 0.92 & -14.53 & 1.53 \\
B5 & 1.35 & 5.55 & 3.81 & 71.39 & 83.60 & 1.72 & 0.80 & -8.58 & 0.84 & -14.24 & 1.50 \\
B6 & 1.40 & 5.50 & 3.64 & 71.04 & 83.15 & 1.73 & 0.80 & -8.62 & 0.84 & -14.11 & 1.50 \\
\midrule
C1 & 2 & 4.74 & 3.02 & 74.93 & 55.25 & 1.94 & 0.84 & 0.69 & 0.81 & -6.33 & 1.76 \\
C2 & 4 & 3.53 & 1.90 & 75.90 & 70.65 & 2.28 & 0.88 & 3.46 & 0.72 & -2.33 & 1.69 \\
\midrule
D1 & 2 & 3.64 & 2.57 & \textbf{75.97} & 71.10 & 2.43 & 0.90 & 7.05 & 0.72 & 0.79 & 1.58 \\
D2 & 7 & 3.19 & 1.53 & 75.49 & 86.55 & 3.53 & 0.95 & 12.56 & 0.58 & 7.18 & \textbf{0.84} \\
D3 & 7.5 & \textbf{3.07} & \textbf{1.49} & 75.28 & \textbf{88.00} & \textbf{3.58} & \textbf{0.96} & \textbf{12.98} & \textbf{0.56} & \textbf{7.33} & 0.89 \\
\midrule
E & 4 & 4.22 & 2.71 & 72.85 & 66.60 & 2.09 & 0.86 & 1.85 & 0.79 & -1.61 & 1.76 \\
\bottomrule
\end{tabular}
}
\label{tab:open_set}
\end{table*}

\begin{figure*}[h]
  \centering
  \subfigure[Automatic speech recognition]{\includegraphics[width=0.46\textwidth]{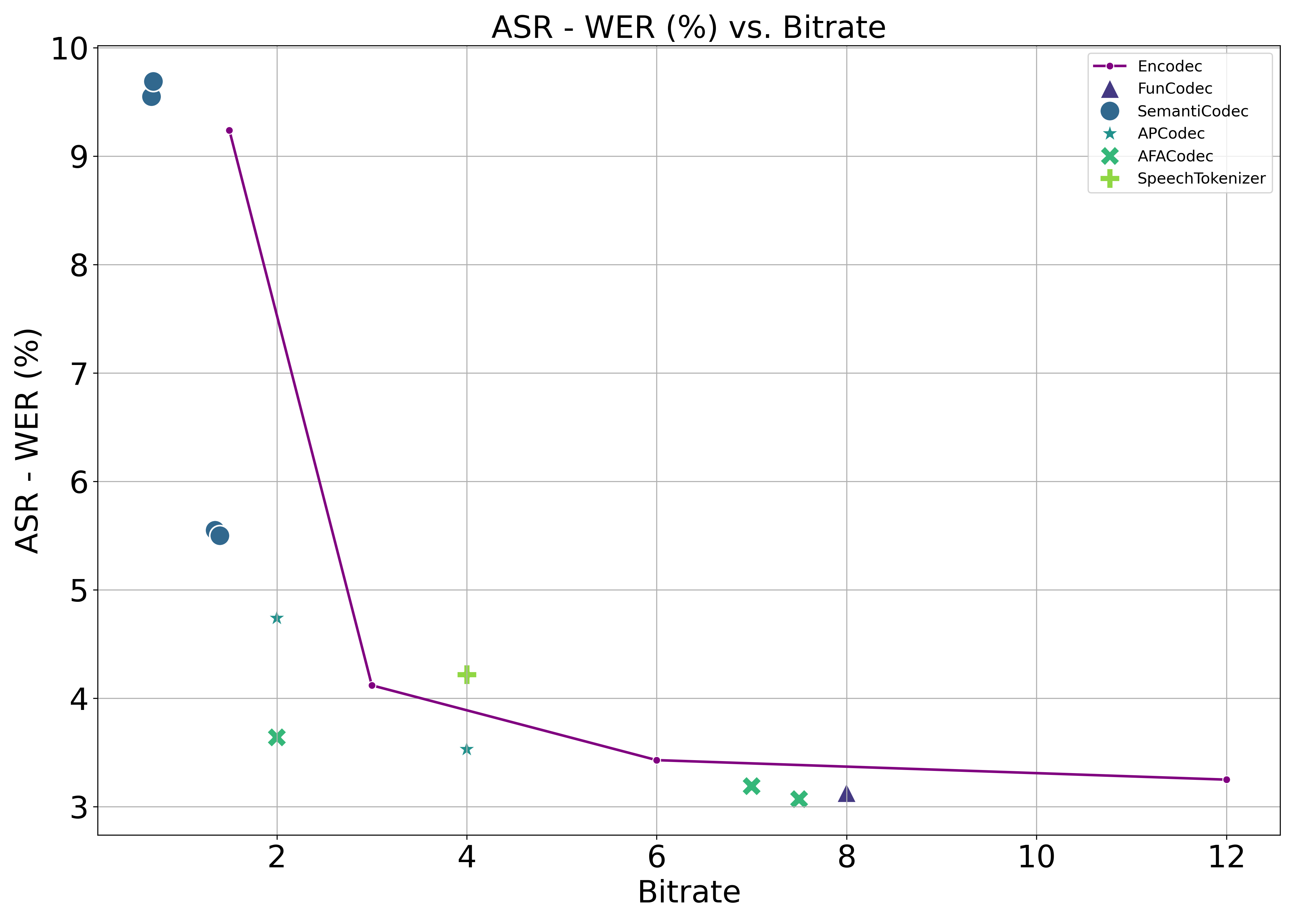}}\hfill
  \subfigure[Automatic speaker verification]{\includegraphics[width=0.46\textwidth]{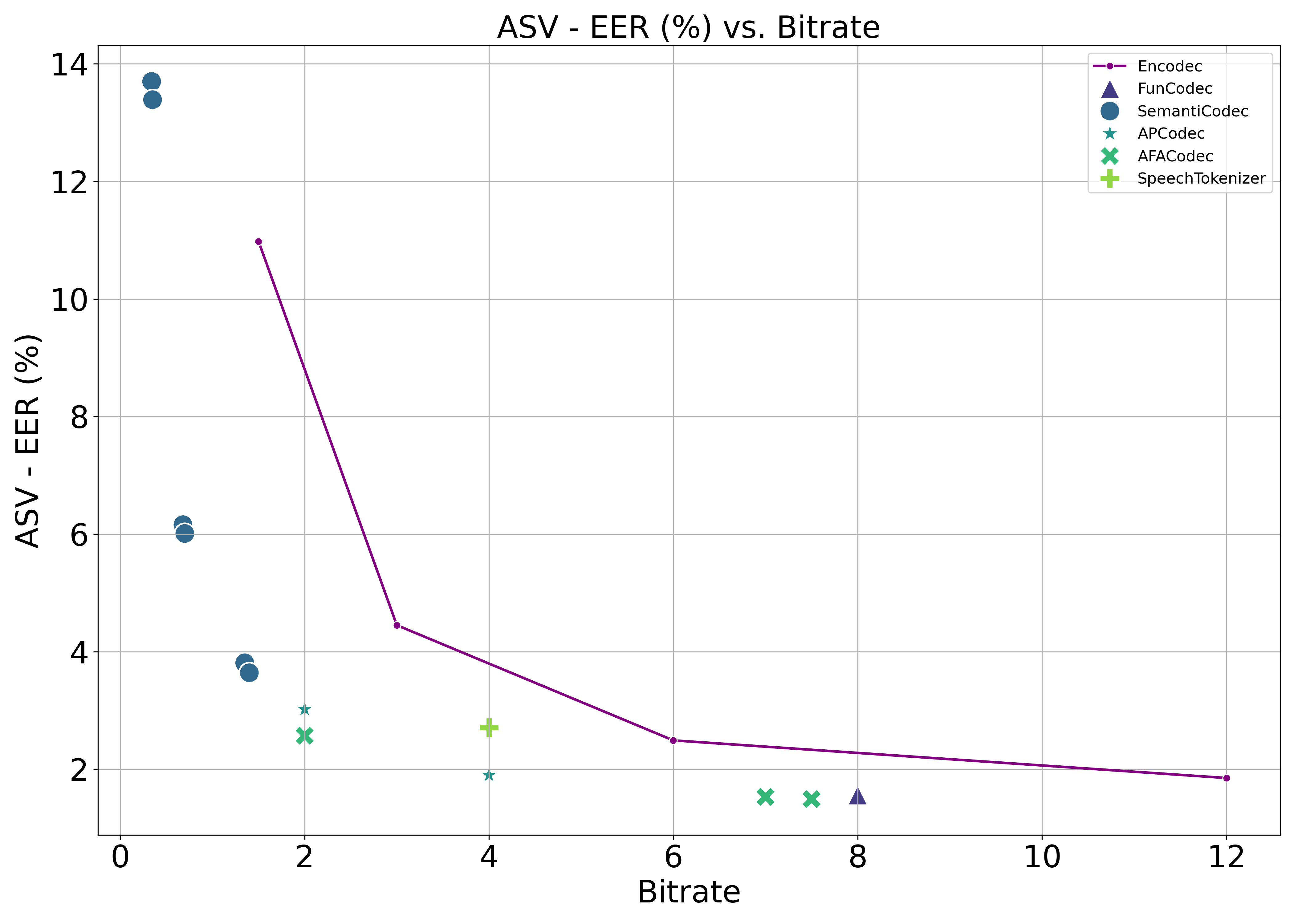}}\\
  \subfigure[Emotion recognition]{\includegraphics[width=0.46\textwidth]{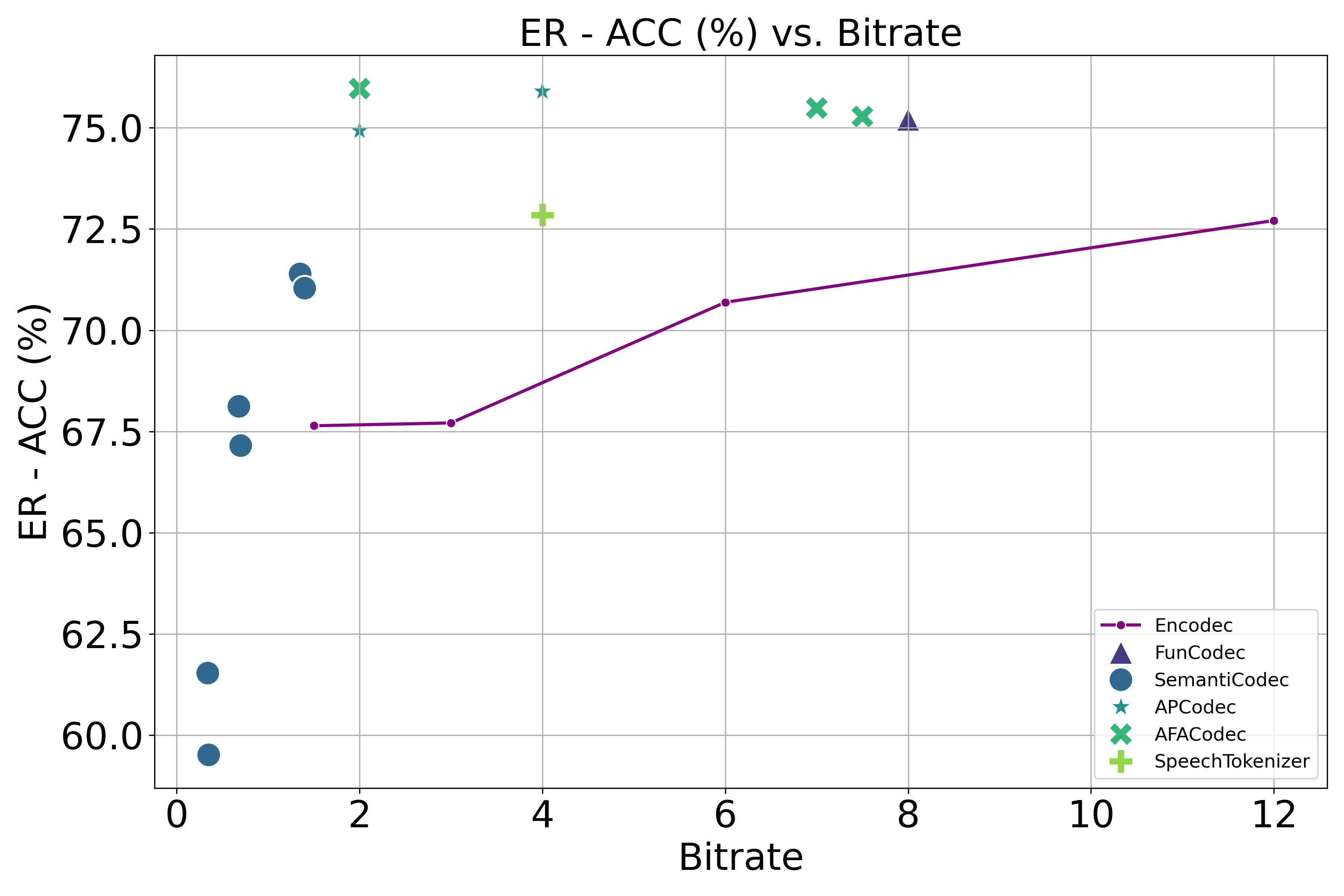}}\hfill
  \subfigure[Audio event classification]{\includegraphics[width=0.46\textwidth]{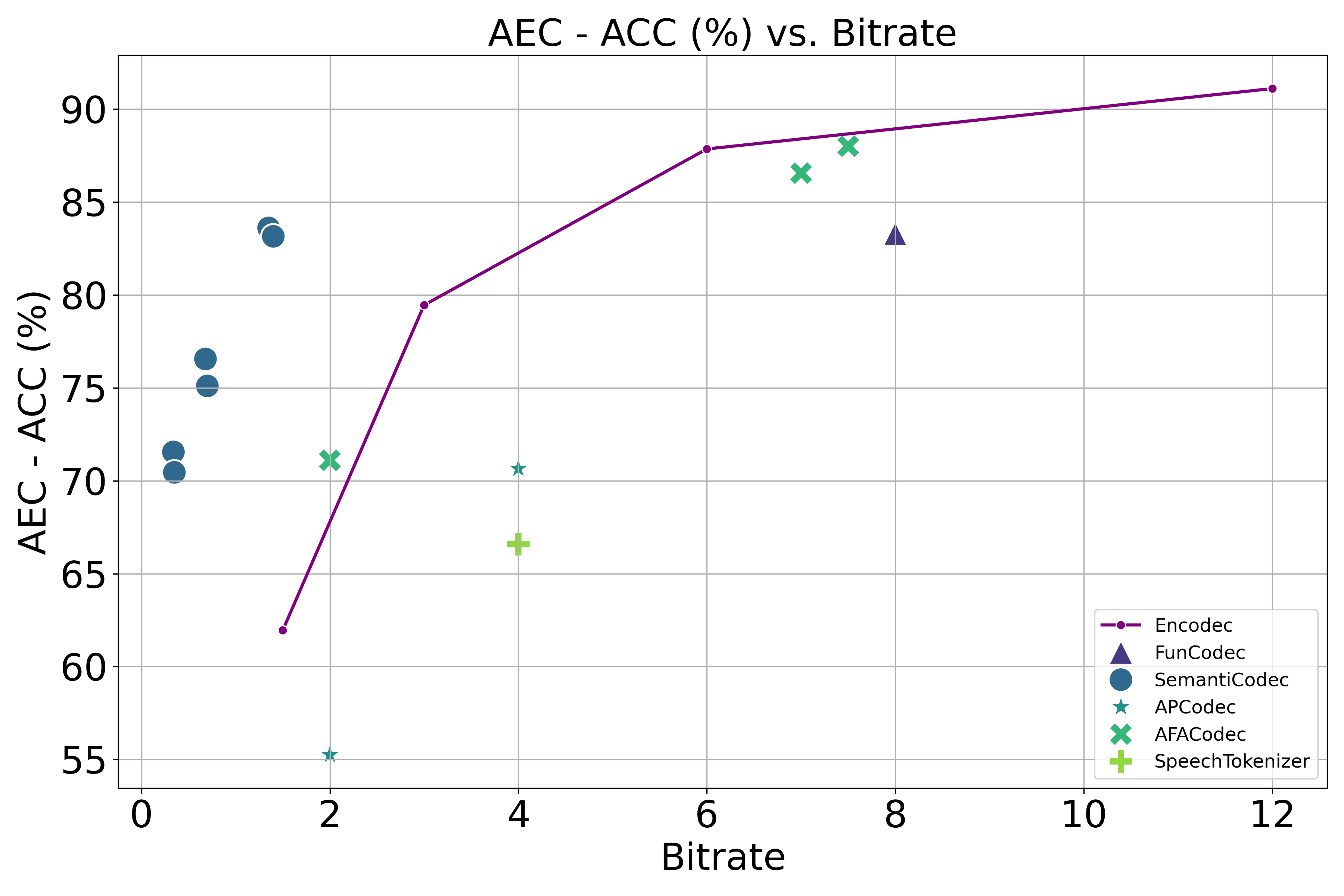}}
  \caption{The application-level evaluation for the open set. We incorporate Encodec \cite{defossez2022high} as the reference for comparison.}
  \label{fig:four_subfigures}
\end{figure*}

\begin{table*}[h]
\centering
\caption{Comparison between codec models for the hidden set. "None" means that no codec has been applied.}
\resizebox{\textwidth}{!}{
\begin{tabular}{llcccccccccccc}
\toprule
\multirow{2}{*}{\textbf{Codec}} & \multirow{2}{*}{\textbf{Bitrate}} & \multicolumn{4}{c}{\textbf{Application}} & \multicolumn{4}{c}{\textbf{Speech Signal-Level Metrics}} & \multicolumn{2}{c}{\textbf{Audio Signal-Level Metrics}} \\
\cmidrule(lr){3-6} \cmidrule(lr){7-10} \cmidrule(lr){11-12}
 & \textbf{(kbps)} & \textbf{WER (\%)} & \textbf{EER (\%)} & \textbf{ACC (\%)} & \textbf{ACC (\%)} & \multirow{2}{*}{\textbf{PESQ}} & \multirow{2}{*}{\textbf{STOI}} & \multirow{2}{*}{\textbf{SDR}} & \textbf{Mel} & \multirow{2}{*}{\textbf{SDR}} & \textbf{Mel} \\
 &  & \textbf{$\downarrow$ (ASR)} & \textbf{$\downarrow$ (ASV)} & \textbf{$\uparrow$ (ER)} & \textbf{$\uparrow$ (AEC)} &  &  &  & \textbf{Loss} &  & \textbf{Loss} \\
\midrule
None & - & 5.28 & 1.60 & 59.60 & 78.01 & - & - & - & - & - & - \\
\midrule
A & 8 & 5.49 & 2.20 & 46.46 & 69.88 & 3.29 & 0.96 & 8.05 & 1.84 & -3.98 & 2.48 \\
\midrule
B1 & 0.34 & 34.95 & 8.20 & 47.47 & 60.69 & 1.49 & 0.76 & -9.15 & 1.21 & -14.36 & 2.09 \\
B2 & 0.35 & 32.97 & 8.40 & 49.49 & 59.70 & 1.50 & 0.76 & -9.14 & 1.20 & -14.29 & 2.08 \\
B3 & 0.68 & 11.09 & 5.00 & 53.54 & 65.94 & 1.82 & 0.82 & -8.07 & 1.00 & -13.22 & 1.94 \\
B4 & 0.70 & 10.34 & 4.80 & \textbf{58.59} & 65.70 & 1.84 & 0.83 & -8.09 & 1.00 & -13.21 & 1.93 \\
B5 & 1.35 & 7.37 & 3.20 & 49.49 & 69.76 & 2.09 & 0.86 & -7.67 & 0.90 & -12.54 & 1.88 \\
B6 & 1.40 & 7.12 & 3.20 & 54.55 & 70.37 & 2.11 & 0.86 & -7.59 & 0.89 & -12.56 & 1.87 \\
\midrule
C1 & 2 & 6.70 & 3.60 & 52.53 & 51.17 & 2.41 & 0.89 & 2.27 & 0.86 & -6.68 & 1.10 \\
C2 & 4 & 6.02 & \textbf{2.00} & 49.49 & 60.77 & 2.83 & 0.93 & 4.50 & 0.77 & -3.09 & 0.98 \\
\midrule
D1 & 2 & 6.01 & 3.00 & 47.47 & 73.89 & 2.89 & 0.93 & 7.88 & 0.77 & -0.36 & 1.86 \\
D2 & 7 & 5.47 & 2.20 & 52.53 & 74.93 & 3.83 & \textbf{0.97} & 13.19 & \textbf{0.61} & 6.56 & 0.96 \\
D3 & 7.5 & \textbf{5.39} & \textbf{2.00} & 50.51 & \textbf{76.00} & \textbf{3.88} & \textbf{0.97} & \textbf{13.66} & 0.62 & \textbf{7.04} & \textbf{0.93} \\
\midrule
E & 4 & 6.04 & 2.60 & 50.51 & 57.82 & 2.68 & 0.91 & 3.66 & 0.79 & -5.36 & 1.95 \\
\bottomrule
\end{tabular}
}
\label{tab:hidden_set}
\end{table*}

\noindent\textbf{VoxCeleb1} \cite{DBLP:conf/interspeech/NagraniCZ17} is an audio-visual dataset featuring short segments of human speech sourced from interview videos on YouTube. We use the test-O set for evaluation.

\noindent\textbf{Libri2Mix}  \cite{cosentino2020librimix} is a synthesized corpus that blends speech from two speakers with background noise sourced from the WHAM! dataset. The speech segments are extracted from LibriSpeech and organized into four subsets: train-360, train-100, dev, and test, totaling 300 hours of speech. We utilize the test set for codec evaluation.

\noindent\textbf{RAVDESS} \cite{livingstone2018ryerson} The Ryerson Audio-Visual Database of Emotional Speech and Song (RAVDESS) is a well-known emotional dataset, licensed under CC BY-NC-SA 4.0. It features performances by 24 professional actors (12 female, 12 male) with North American accents. The dataset includes speech expressing calm, happiness, sadness, anger, fear, surprise, and disgust.

\noindent\textbf{CREMA-D} \cite{6849440} comprises 7,442 clips performed by 91 actors (48 male and 43 female), with each clip annotated for six distinct emotions. These professional actors, guided by experienced theatre directors, skillfully express designated emotions while delivering specific sentences

\noindent\textbf{VoxLingua107 Top 10} \cite{valk2021slt} contains audio segments designed for spoken language identification, covering 107 distinct languages. The dataset consists of audio clips automatically extracted from YouTube videos, focusing on the top 10 most frequent languages.

\noindent\textbf{ESC-50} \cite{piczak2015dataset} comprises 2000 environmental sounds categorized into 50 classes. These clips are manually selected from public field recordings compiled by the Freesound.org project.

\noindent\textbf{FSD50K} \cite{fonseca2022FSD50K} A is an open collection of human-labeled sound events, consisting of 51,197 Freesound clips categorized into 200 classes from the AudioSet Ontology. For codec evaluation, we utilize the test and validation sets.

\noindent\textbf{Gunshot Triangulation} \cite{cooper2020gunshots}  captures audio recordings of seven distinct firearms—four pistols and three rifles—each fired at least three times. The shots were aimed sequentially at a target positioned 45 meters away from the shooter in an open field. The sound of these firings was captured using four separate iPod Touch devices.

\subsubsection{Hidden set}
We have created a hidden set comprising counterparts for all types of datasets in the open set. To construct these hidden datasets, we collaborated with LxT\footnote{https://www.lxt.ai/} to engage 60 human speakers, ensuring gender balance, to recite sentences and record the audio.

\section{Submissions}
\label{sec:Submissions}


In total, this challenge received 5 submitted codec models with different setups, resulting in 13 distinct settings as illustrated in Table~\ref{tab:codec_info}.
We will assess their performance on both the open and hidden sets. 
Additionally, the Encodec \cite{defossez2022high} serves as a reference for comparison. 
We have also included codecs developed by ESPnet-Codec\footnote{https://github.com/espnet/espnet/tree/codec} (under controlled settings) for further analysis using our evaluation pipeline.

\subsection{Overview of submitted codec models}

\textbf{FunCodec (A) \cite{du2023funcodec}}: Unlike many codec models that concentrate on the time domain, FunCodec introduces a frequency-domain approach. The authors assert achieving comparable performance with fewer parameters and lower computational complexity. Additionally, they find that integrating semantic information into the codec tokens enhances speech quality at low bit rates.

\textbf{SemantiCodec (B) \cite{liu2024semanticodec}}: SemantiCodec leverages a dual-encoder architecture: a non-trainable semantic encoder to capture the main semantic information from audio and a trainable acoustic encoder to capture detailed residual information. The semantic encoder uses a self-supervised AudioMAE \cite{huang2022masked} to extract features, followed by k-means clustering to generate semantic codes. The input features, along with the quantized embeddings from the semantic encoder, are then fed into the trainable acoustic encoder to capture the remaining details and produce acoustic codes. Their experiments demonstrate that their semantic codes provide rich information for audio event classification and understanding, even at remarkably low bitrates (0.47 kbps).

\textbf{APCodec (C) \cite{ai2024apcodec}}: APCodec delivers high-quality audio at a low bitrate with fast generation speed and low latency, specifically designed for 48 kHz audio. Unlike other recent codec models, APCodec encodes and decodes both amplitude and phase spectra. To make the model causal without losing performance, a non-causal teacher model is used to train the streamable APCodec through knowledge distillation.

\textbf{AFACodec (D)}: A Neural Speech Codec with Plug-and-Play Adaptive Feature Awareness. 
AFACodec's training framework is based on the descript-audio-codec with several modifications. It includes an encoder that converts the time-domain waveform into latent features and an RVQ quantizer that quantizes these latent features into code vectors within a codebook. The codebook size is 1024, and the code vector dimension is 8. Additionally, a decoder reconstructs the waveform from the quantized features. Notably, a dual-stream adaptive feature-aware module has been introduced before and after quantization. This plug-and-play module focuses on identifying the most important positions and features from both temporal and channel dimensions, thereby enhancing coding efficiency and reducing redundancy.

\textbf{SpeechTokenizer (E) \cite{zhang2023speechtokenizer}}: SpeechTokenizer is a unified speech tokenizer tailored for speech-language models. It adopts an Encoder-Decoder architecture enhanced with RVQ. By incorporating semantic and acoustic tokens, SpeechTokenizer hierarchically separates different aspects of speech information across multiple RVQ layers. Specifically, the first RVQ layer is designed to regularize learning of the Hubert tokens \cite{hsu2021hubert}. The authors argue that these techniques enhance the disentanglement of information across RVQ layers.

\subsection{Results and analysis}

We present the challenge results in Table~\ref{tab:open_set} (open set; visible to participants) and Table~\ref{tab:hidden_set} (hidden set).
Figure~\ref{fig:four_subfigures} also provided an overview (for the open set) of the trade-off between bitrate and application performance, with the well-known pioneer codec model Encodec~\cite{defossez2022high} serving as the baseline.

\subsubsection{Open set}

By comparing the codec models in Table~\ref{tab:open_set} and Figure~\ref{fig:four_subfigures}, we have the below observations:
\begin{itemize}
    \item In the mid-bitrate range ($7\sim 8$ kbps), AFACodec (D) consistently stands out as the best model for speech applications. It achieves the lowest Word Error Rate of 3.07\% for ASR, the lowest Equal Error Rate of 1.49\% for ASV, approximately 75\% accuracy in emotion recognition—nearly matching the original audio's 76.76\% with less than a 1\% relative performance drop, and the highest accuracy of 88\% for AEC.
    \item At $4$kbps, APCodec (C) outperformed SpeechTokenizer (E) on almost all metrics, including ASR WER, despite SpeechTokenizer being designed specifically to encode the semantic content of speech.
    \item However, all participating models failed to beat the baseline model Encodec in the $4\sim 8$ kpbs range on audio event classification.
    \item In the low-bitrate range ($\leq$ 2 kbps), SemantiCodec delivered strong results. It is worth noting that SemantiCodec outperformed the baseline model and other models by a significant margin in audio event classification (Figure~\ref{fig:four_subfigures}(d)), demonstrating excellent audio information preservation as claimed in the authors' paper \cite{liu2024semanticodec}.
\end{itemize}

\begin{table}[h]
\centering
\caption{Correlation Matrix between applications and objective metrics. Correlation scores close to -1 indicate a strong negative correlation, while correlation scores close to 1 indicate a strong positive correlation.}
\label{tab:correlation}
\resizebox{0.4\textwidth}{!}{%
\begin{tabular}{ccccc}
\toprule
\textbf{} & PESQ & STOI & SDR & Mel Loss \\
\hline
ASR-WER & -0.588 & \textbf{-0.807} & 0.272 & -0.595 \\
ASV-EER & -0.696 & \textbf{-0.891} & -0.712 & 0.254 \\
ER-ACC & 0.732 & \textbf{0.920} & 0.793 & -0.262 \\
AEC-ACC & - & - & 0.233 & -0.497 \\
\bottomrule
\end{tabular}%
}
\end{table}

\subsubsection{Hidden set}

Hidden and open sets have similar trends. From Table~\ref{tab:hidden_set}, we have the following observations:
\begin{itemize}
    \item AFACodec (D) excels in the mid-bitrate range ($7\sim 8$ kbps). Notably, D1 at 2 kbps outperforms both E and C2 at 4 kbps in audio event classification and performs comparably in ASR and ER.
    \item Even at very low bitrates, SemantiCodec (B) performs very well in audio event classification, e.g. B6 achieves 70.37\% with only 1.4 kbps.
\end{itemize}

\subsubsection{Correlation analysis}
The correlation matrix for the open set as in Table~\ref{tab:correlation} shows the Pearson correlation coefficients between applications (ASR-WER, ASV-EER, ER-ACC, AEC-ACC), and the metrics (PESQ, STOI, SDR, and Mel Loss). For audio datasets, we only calculate SDR and Mel Loss. Key observations about the speech application level metrics are:
\begin{itemize}
    \item The STOI demonstrates strong negative or positive correlations across the three tasks (ASR, ASV, ER), with correlation scores less than -0.8 for ASR-WER and ASV-EER and a correlation score of 0.92 for ER-ACC. This indicates that speech intelligibility strongly influences performance in these applications.
    \item PESQ serves as the second reliable metric, showing comparably balanced correlations with ASR-WER, ASV-EER, and ER-ACC. SDR also shows reasonable correlation scores with ASV-EER and ER-ACC.
    \item In contrast, Mel Loss exhibits very weak correlations: e.g., 0.254 for ASV-EER and -0.262 for ER-ACC.
\end{itemize}

As shown in the fourth row of Table~\ref{tab:correlation}, regarding the audio-related metrics, the Pearson correlation coefficients between AEC-ACC and SDR, and between AEC-ACC and Mel Loss, are only 0.233 and -0.497, respectively. 
This indicates that the accuracy of audio event classification has a weak correlation with objective metrics such as SDR and Mel Loss. This may be attributed to certain codec models (e.g., SemantiCodec) employing generative models like diffusion models \cite{ho2020denoising} as decoders to reconstruct the audio signals, potentially causing time shifts between audio samples.

\subsection{Take-away}
Finally, we want to summarize and conclude the results for submissions with the following takeaways:
(1). This challenge highlighted that prevalent codec models like Encodec are not perfect, particularly when the bitrate decreases to very low levels, such as 2 kbps. 
(2). In the mid-bitrate range ($7\sim 8$kbps), AFACodec outperformed other neural codec models with strong results on speech applications.
(3). SementiCodec showed that a specialized codec model can significantly reduce bitrate with a minimum information loss on audio applications.
It is a suitable codec model for downstream tasks that favor low-bitrate inputs (e.g., audio large language models).
(4). STOI is a comparably reliable metric for speech downstream applications than the other three metrics. However, Mel Loss and SDR exhibit weak correlations with audio event classification accuracy.


\section{Conclusions}
\label{sec:conclusions}

This paper reviews five models participating in this challenge, present our findings, and highlight insights into codec models.
Encodec struggles at very low bitrates (e.g., 2 kbps).
AFACodec performs best at mid-range bitrates (7-8 kbps), especially for speech.
SementiCodec effectively reduces bitrate with minimal audio information loss.
We also provide a training-free and computationally efficient evaluation pipeline to help codec developers quickly obtain preliminary results and gain intuitions, which can serve as a reference for further development.
In the future, we plan to include multi-lingual datasets for evaluation.

\ninept
\bibliographystyle{IEEEbib}
\bibliography{strings,refs}

\end{document}